\documentclass[preprint,aps,showpacs]{revtex4}  
\usepackage{graphicx}%
\newcommand{\figwidth}{6in}
\begin{document}


\title{A quantum-dot array as model for Copper-Oxide superconductors:
A dedicated quantum  simulator for  the  many-fermion problem}
\author{Efstratios  Manousakis} \affiliation{  Department  of Physics  and
Center  for   Materials  Research  and   Technology,\\  Florida  State
University, Tallahassee, FL 32306-4350 } \date{\today} 
\begin{abstract}
Quantum systems  with a large  number of fermionic degrees  of freedom
are intractable by quantum simulations. In this paper we introduce the
concept of  a dedicated quantum simulator(DQS) which  is an artificial
system of quantum dots whose  Hamiltonian maps exactly to the original
many  fermion problem.  While  the universal  quantum simulator  (UQS)
introduced  by Feynman  in 1982  can simulate  any  quantum mechanical
many-body  problem,  a DQS  can  only  solve  a particular  many  body
problem.  Our concept of the dedicated quantum simulator is not a 
quantum computer but rather a quantum ``analog'' device, dedicated to a
particular quantum computation. As an example, we consider  the system 
of the $CuO$ plane in the copper-oxide superconductors and we propose  
an array of electrostatically confined quantum dots to be used as its 
dedicated quantum simulator.  We show that  this dedicated device can 
be used to image stripe formation as a function of the electron doping 
using electric force microscopy. We argue that such a dedicated quantum 
simulator may be easier to realize in the future compared to a general
purpose quantum computer.
\end{abstract}
\maketitle
\section{Introduction}
In  simulations  of  quantum  many-fermion systems  large  statistical
fluctuations  arise  due to  cancellations  among  large amplitudes  of
configurations differing  by fermion exchanges. As  a consequence, the
computational  time required to  obtain acceptable  statistical errors
grows exponentially  with the system  size. This limitation  is severe
because only small-size systems can  be simulated and that prevents us
from being able  to extrapolate to the thermodynamic  limit. This is a
general  problem  in  several  fields  of  computational  physics  and
chemistry and  it would  be of general  importance if  a computational
instrument   that   simulates  quantum   fermion   systems  could   be
constructed.

Progress  in quantum  computation\cite{divincenzo1}  has raised  hopes
that  a  realistic  computation  using  a quantum  computer  could  be
achieved in the future. Feynman had conjectured\cite{feynman} that the
quantum  computer   can  be  used   to  simulate  any   local  quantum
system. Later, Lloyd showed\cite{lloyd} that a quantum computer can be
programmed so that it can be such a universal quantum simulator (UQS).
Recently  DiVincenzo  et al.\cite{divincenzo3}  have  proposed that  a
coupled  quantum-dot pair may  be used  to represent  a q-bit.  However,
in any attempt to design a general purpose quantum computer one needs to find
an  approach to  externally manipulate  quantum mechanical  states, to
preserve the  quantum coherence of these  states for some  time and to
transport  them at  a  macroscopic distance  away  before the  quantum
information is dissipated.

In  this  paper  we  introduce  the concept  of  a  dedicated  quantum
simulator (DQS) in  contrast to the UQS.  It will  become clear that a
dedicated quantum  computer to simulate a  specific many-fermion model
does not require a controlled  initial state or building quantum gates
which also need  to be controlled. A  DQS as defined in this paper, 
while it should be a system without significant amount of 
impurities or other defects which could create decoherence, it does 
not require from us to manipulate  q-bits where destruction of 
coherence can also occur. Thus, building a DQS is a more realistic 
goal in comparison to building a universal quantum computer.

Our concept of the dedicated quantum simulator is not a quantum 
computer but rather a quantum ``analog'' device, dedicated to a 
particular quantum computation. Long time ago,
before  digital classical computers became fast enough to 
carry out numerical integration or differentiation, the so-called 
analog computers were used for that purpose. To obtain the 
integral or the derivative of a function $f(t)$, an electrical 
time-dependent current, which changes in time in the same fashion 
as the function $f(t)$, was used as input to a circuit 
which contains a capacitor or an impedance. For the integral 
or the derivative of the input function one would measure the 
voltage across the capacitor or across the impedance respectively.
No digital computation was carried out by such a device, but the
entire ``computation'' was based on the physical property of the
device used. A particular physical circuit was dedicated to
a specific computation, i.e., the capacitor circuit for 
integration, while the  impedance circuit for differentiation.
In this paper we introduce a dedicated quantum simulator
whose relationship to a quantum computer is analogous to the
relationship of a digital classical computer to  an
analog classical computer. Notice that we are careful and we use the
term ``simulator'' as opposed to the term ``computer''.

It is well known in statistical mechanics that the Landau-Ginzburg 
model of a superfluid can be mapped onto the  $X-Y$ model.
The $X-Y$ model can also describe the critical fluctuations of
certain types of magnetic systems where the order parameter 
is a two component vector. Thus, instead of using a digital
computer to compute the critical exponents associated with the
superfluid to normal-fluid phase transition, one can study the 
experimental results obtained on such a designed magnetic system
assuming that the connection (the mapping) of the 
magnetic system to the $X-Y$ model is accurate.
One could therefore think of this model magnetic system as
a {\it dedicated simulator} of the critical properties of
the superfluid. 

The idea can be extended further to a pure quantum many body system
where we are interested in the statistical properties of that system.
If we  could prepare a physical  system which is described  by a known
model quantum Hamiltonian, all we would  have to do is perform 
measurements of the desired observables. It is rather hopeless to 
expect that we could configure  atoms together to  interact in  our 
desired  way as  in the model Hamiltonian.  Quantum dots share  
many features with  the atomic spectra  and  they are  sometimes  
called  ``artificial'' atoms.   The parameters defining  a quantum 
dot can be  artificially controlled and designed. In addition, 
one can create arrays of such dots where we can manipulate their 
interactions.  Therefore, if we could design an array
of quantum  dots interacting in  a similar way, the  original physical
system can exist  at a very different energy scale but  as long as the
model system (used in the simulation) shares the same geometry and the
same  values of  doping  and dimensionless  parameter  ratios one  can
directly compare dimensionless ratios of observables using scaling.
This is what we call ``dedicated quantum simulator'' and 
 has nothing to do with the functions involved in quantum computing,
just in the same way as the question of how a classical digital 
computer works is irrelevant to the problem of an analog computation.

Making  a   quantum  computer  that  performs   operations  which  are
controllable at the so-called q-bit level is a far more difficult task
than making a device that can perform such a dedicated task to solve a
specific quantum many fermion Hamiltonian. The  reason is that  every nature's
operation is quantum mechanical and  thus we can ``take advantage'' of
that  and instead  of breaking  down the  problem into  a huge  set of
classical operations  we prepare a  many-body quantum system  which is
described by,  and thus can  mimic exactly, the  theoretical many-body
Hamiltonian which we wish to solve.

In  this paper,  we choose  to give  an example  of  a two-dimensional
quantum-dot array  which can be  mapped to a  Hubbard-like Hamiltonian
identical to that used to describe  the physics of the $Cu-O$ plane of
the  high  temperature  superconductors.  This serves as model
for the copper-oxide plane in the copper-oxide superconductors
based on a quantum dot two-dimensional array.
The  Hamiltonian which describes the quantum-dot array does  not
contain phonons as degrees of  freedom and thus, one can determine the
physical properties in the  absence of phonons. The quantum-dot system
exists  at an  energy scale  (a few  $meV$) which  is three  orders of
magnitude smaller than that of  real physical system. We think of this
system of  the array  of quantum  dots as a  quantum simulator  of the
physics of the original system. We also discuss that this system
should form stripes at the appropriate filling factor in an analogous 
manner to that in the copper-oxide superconductors. In addition, we  
discuss how to use this model to study the formation of 
stripes in the original problem of the copper-oxide planes.

\section{The Quantum Dot Array}

We wish to consider a  two-dimensional electron gas (2DEG) which forms
in  an $Al_xGa_{1-x}As/GaAs$  heterostructure. Such  a heterostructure
can be grown  by molecular beam epitaxy (MBE) on  a $n^+$ doped $GaAs$
substrate.   On  top  of  this   layer  one  grows  a  layer  of  pure
$AlGaAs$. Next, a layer of pure  $GaAs$ is grown which has smaller gap
than  $AlGaAs$.  The  2DEG is  formed  in  this  last layer  near  the
interface with the $AlGaAs$ layer.   A positive voltage applied to the
$n^+$    doped    substrate    controls    the    density    of    the
2DEG.   Two-dimensional   electron   densities   of   the   order   of
$n=10^{11}/cm^{-2}$  are desirable  for the  application  described in
this paper.   A spacer  of pure $GaAs$  between the substrate  and the
$AlGaAs$ material  may also be  necessary to increase the  mobility of
the electrons at the interface.

\begin{figure}[btp]
\includegraphics[width=\figwidth]{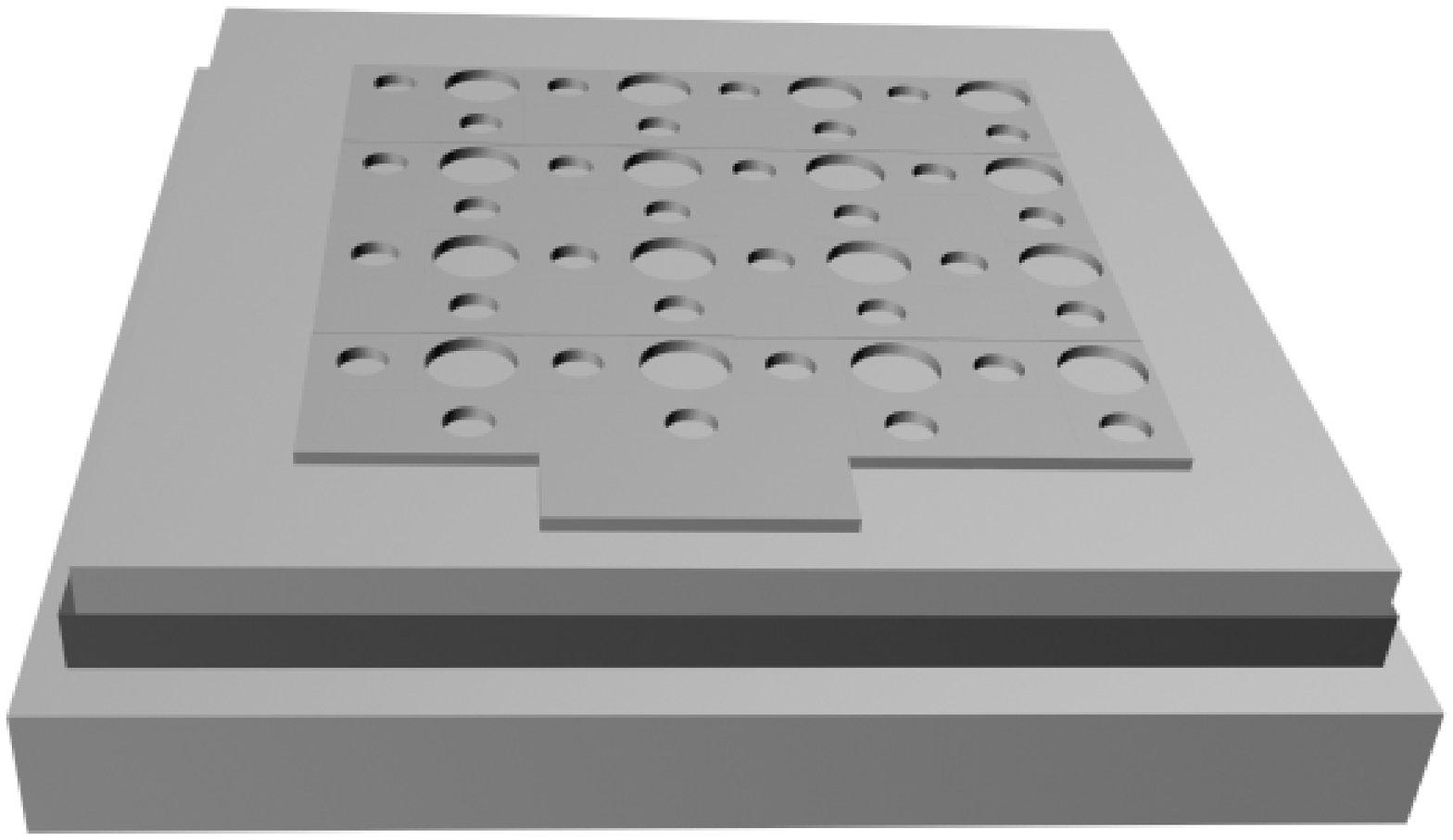}
\caption{Quantum dot  model for the  $Cu-O$ plane of  the copper-oxide
superconductors.}
\label{hetero2}
\end{figure}

In  this  paper, we  consider  the  case of  the  $CuO$  plane of  the
copper-oxide  superconductors.   For  this  example,  we  propose  the
metallic gate with the hole pattern shown in Fig.~\ref{hetero2}, i.e.,
with  an   array  of  two   different  size  holes  placed   onto  the
heterostructure  as   the  top   electrode.   Such  patterns   can  be
``drilled''  on  a thin  metallic  plate  with  e-beam lithography.  A
negative gate voltage $V_g$ is applied between this gate and the 2DEG.

\begin{figure}[btp]
\includegraphics[width=\figwidth]{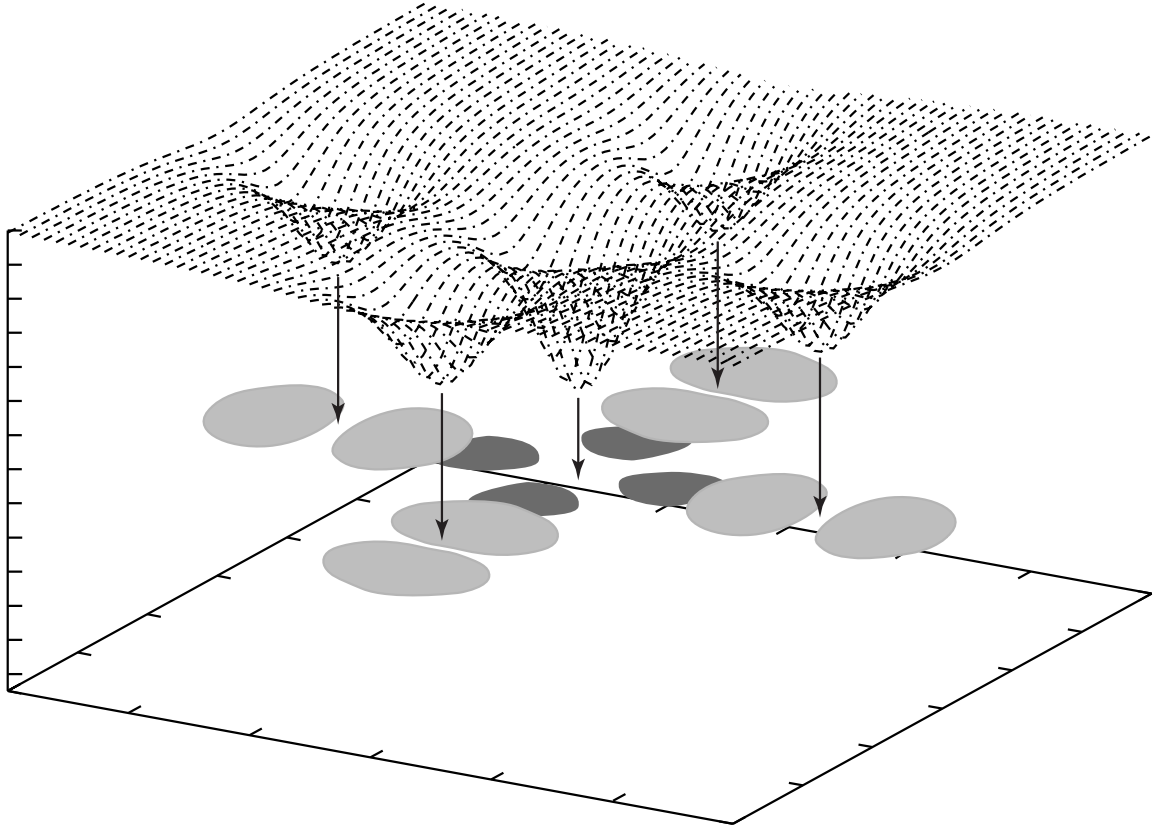}
\caption{The potential  landscape for a  cluster of quantum  dots with
the central dot somewhat larger  than the other four surrounding dots.
Below the potential minima the  orbitals of the outer electrons, i.e.,
the $d_{x^2-y^2}$  of the central dot  and the $p_x$ and  $p_y$ of the
surrounding  smaller  dots are  shown.  Inside  the  shaded areas  the
magnitude of the wavefunction is larger than half its peak value.}
\label{3dpot}
\end{figure}

First, let  us consider  a single  hole of radius  $a$ created  on the
metallic  gate. At  a distance  $d$ from  the gate,  the  potential is
modified  from its  value in  the  absence of  the hole  by an  amount
$\delta V_a(\vec r)$ given by\cite{jackson}
\begin{equation} 
\delta  V_a(\vec  r)=  -{{E   a^2}  \over  {\pi}}  \int_0^{\infty}  dk
j_1(ka)e^{-k d} J_0(k r).
\end{equation} 
where  $E=V_{g}/(\epsilon d)$  is  the electric  field  below the  top
metallic  gate  in the  absence  of the  hole  and  $\epsilon$ is  the
dielectric  constant  for pure  $GaAs$.   Here  we  have considered  a
cylindrical coordinate  system with the $z$ axis  perpendicular to the
plane and passing through the center of the hole with $r = \sqrt{x^2 +
y^2}$ the distance from the axis.  The electrons are thus trapped a in
quantum well  and as they move  away from $r=0$ they  feel a parabolic
repulsion which for small $r$ is given by
\begin{eqnarray}
\delta V(\vec  r,a) &= \delta V(0,a)  + {1 \over  2} m^* \omega^2_0(a)
r^2
\label{harmonic}\\
\omega_0(a) &= \sqrt{{|e| E a} \over {\pi m^*} } {a \over {(d^2+a^2)}}
\label{freq}\\ \delta V(0,a) &= {{- |e| E a} \over { \pi}} \Bigl ( 1 -
{d \over a} tan^{-1}({a \over d}) \Bigr ),
\label{depth}
\end{eqnarray}
where $m^*$  is the electron  effective mass.  We consider  a metallic
gate with the array of holes shown in Fig.~\ref{hetero2} which produce
a modification  of the  external field at  the interface which  is the
superposition of the change caused by each hole:
\begin{eqnarray}
V(\vec r) = &\sum_{\vec R,\vec \sigma}  \bigl ( \delta V(\vec r - \vec
 R,a_1) + \delta V(\vec r - \vec R + \vec \sigma, a_2)\bigr )
\end{eqnarray}
where $\vec R =  (n_x \hat x + n_y \hat y)  b$ spans the entire square
lattice of  lattice spacing  $b$ formed by  the centers of  the larger
holes  of radius  $a_1$.  The  smaller holes  of radius  $a_2$  are at
positions $\vec R + \vec  \sigma$ where $\vec \sigma$ takes the values
$b/2 \hat x$ and $b/2 \hat y$.  This potential can be considered as an
external  field felt  by  the 2DEG  and  to which  the electrons  will
respond.  If the negative gate potential is not strong enough to cause
total depletion  of the heterojunction  from the 2DEG,  the electronic
charge  of the  heterojunction  will  move to  benefit  from the  less
repulsive  potential near  the holes.   The  self-consistent potential
landscape on the heterojunction will depend on the 2D electron density
and  is   expected  to  look   schematically  as  the  one   shown  in
Fig.~\ref{3dpot} for  a cluster  of one hole  surrounded by  4 smaller
ones.

We  will  assume  that the  electrons  in  the  dots feel  a  harmonic
oscillator  potential.  Clear  experimental indication\cite{capacitor}
that   electrostatically   confined   quantum  dots   feel   parabolic
confinement comes from  the magic numbers observed.  Let  us denote by
$|n_x,n_y>$ the eigenstates of the 2D harmonic oscillator in Cartesian
coordinates.  In  the case where  we have square lattice  symmetry the
circular symmetry of  the ``atomic'' potential is reduced  and we need
to consider irreducible representations of the group $C_{4v}$.  In our
illustrative example of  the array the dots have  at most 12 electrons
per dot.  In these cases we will need only the following orbitals:

i) The  state with lowest energy  is the state  with $s$-wave symmetry
given  by 
\begin{equation}
\langle \vec  r| s  \rangle =  \langle \vec  r |n_x=0,n_y=0
\rangle = \sqrt{{{\lambda} \over {\pi}}} e^{-{{\lambda r^2} / 2}},
\end{equation}
where $\lambda = m^* \omega /\hbar$.

ii) The  next excited  states are the  two degenerate $p$  states. The
$p_x$ state given by 
\begin{equation}
\langle \vec r| p_x \rangle = \langle \vec r |1,
0 \rangle = {\sqrt{2 \over  {\pi}} \lambda} x  e^{-{{\lambda r^2} / 2}}.
\end{equation} 
The state $|p_y > = | 0,1 >$ obtained by replacing $x$ with $y$.

iii)  The  $d$  wave  states  are  also  separated  according  to  the
representations of the $C_{4v}$.   

a) The state $| d_{x^2+y^2} \rangle
= {1 \over {\sqrt{2}}} (| 2, 0  \rangle + | 0, 2 \rangle )$ belongs to
the representation  $\Delta_1$ and  it is given  by 
\begin{equation}
 \langle  \vec r|
d_{x^2+y^2}  \rangle  =\sqrt{{{   \lambda} \over  {\pi}}}  (1-\lambda  r^2)
e^{{-{{\lambda r^2}/ {2}}}}. 
\end{equation}

 b)  The state $| d_{x^2-y^2} \rangle =
{1 \over {\sqrt{2}}} (|2, 0 \rangle  - | 0, 2 \rangle)$ belongs to the
$\Delta_2$  representation  and  it  is  given  as  
\begin{equation}
\langle  \vec  r|
d_{x^2-y^2}   \rangle  ={{  \lambda^{3/2}} \over   {\sqrt{\pi}}}  (y^2-x^2)
e^{{-{{\lambda r^2}/  2}}}.
\end{equation}

  c)  The state $|d_{xy}  \rangle =  | 1,
1\rangle $ belongs to the $\Delta_{2'}$ representation and it is given
by  
\begin{equation}
\langle \vec  r| d_{xy}  \rangle  = 2  \lambda \sqrt{{\lambda} \over
{\pi}} x y e^{{-{{\lambda r^2} / 2}}}.
\end{equation}

Now we wish to consider the square lattice arrangement of quantum dots
presented in  Fig. ~\ref{hetero2} with  quantum dots of  two different
radii  $a_1$  and  $a_2$  with  $a_1>a_2$.  In  the  smaller  dot  the
potential  starts from  a  higher  value at  its  center. The  spacing
between the energy spectra  depends on the frequencies $\omega_0(a_1)$
and  $\omega_0(a_2)$.   By  controlling  the  relative  depth  of  the
potential in the dots (by changing  $V_g$ and $a_1$ and $a_2$) and the
density of the  2DEG (using the back-gate voltage),  we can create the
following situation.   The larger dot  is filled with 12  electrons in
the $s^2p^4d^6$ pseudo-atomic configuration and the smaller dot with 6
electrons  in  the  $s^2p^4$  pseudo-atomic  configuration.  Thus  the
highest occupied levels are the $d$ for the large dots and the $p$ for
the small dots. Next, we show how this can be achieved.

We need  to estimate  the required quantum-dot  sizes and  2D electron
densities necessary  for producing the case discussed  in the previous
paragraph.  We need to satisfy the following condition:
\begin{equation}
\delta V(0,a_1) + \mu(N_1,a_1) = \delta V(0,a_2) + \mu(N_2,a_2)
\label{equal-mu}
\end{equation}
where $\mu(N,a)=E(N,a)-E(N-1,a)$ is the chemical potential for each of
the quantum  dots in the  presence of only  the quadratic term  of the
interaction  in Eq.~\ref{harmonic}.  Here  $E(N,a)$ is  the total  dot
energy as a function of the electron number and we need to distinguish
$E(N,a_1)$ from $E(N,a_2)$ because of the two different dot sizes.

\section{A Model for a Single Quantum Dot}
\begin{figure}[btp]
\includegraphics[width=\figwidth]{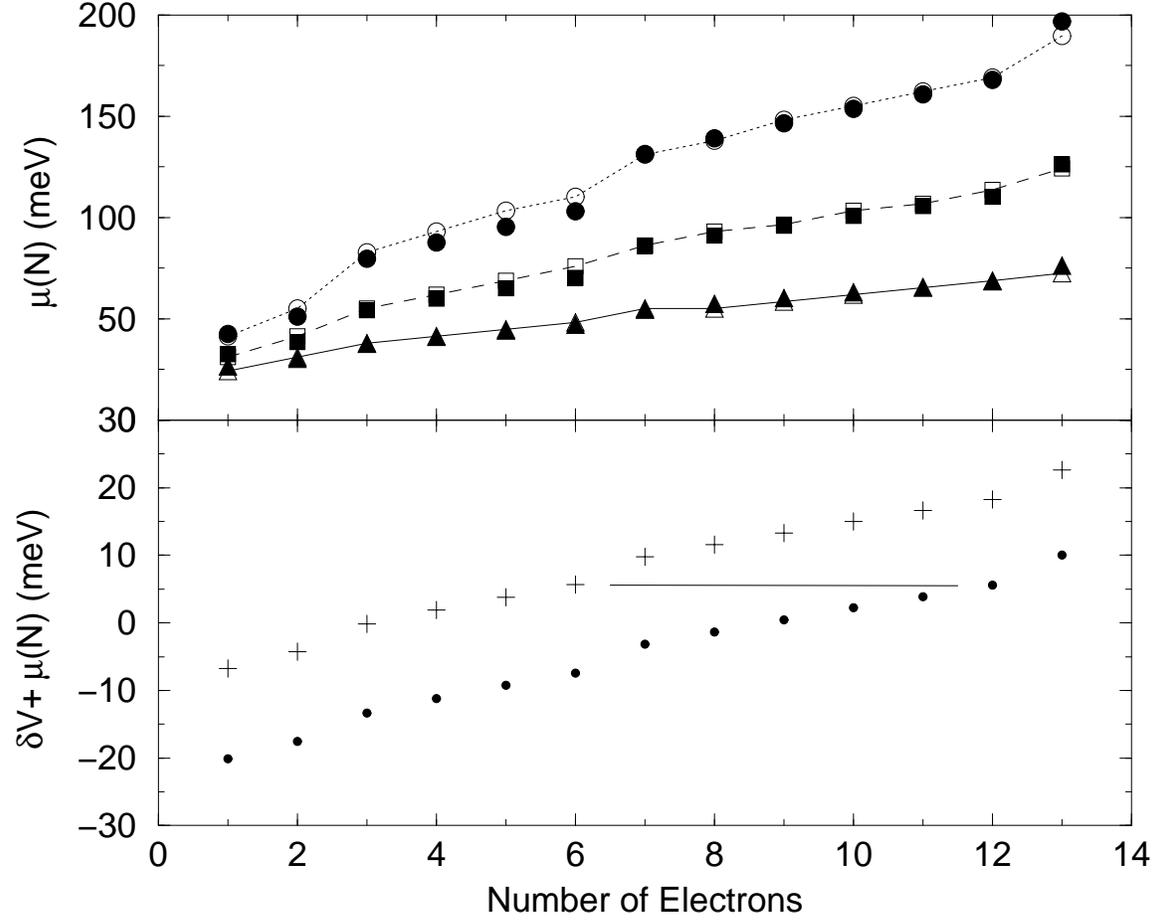}
\caption{Top: The  results of our  model calculation for  quantum dots
(solid  symbols) are compared  with results  of DFT  calculation (open
symbols)  for the  cases of  $\hbar \omega_0=4~meV$  (triangles), $10~
meV$ (squares) and $20~meV$ (circles).  Bottom: The chemical potential
of the two different size dots  match for electron numbers 6 (+ signs)
and 12 (solid circles). }
\label{dft}
\end{figure}

There are several calculations for  a single quantum dot using various
approximations\cite{haw,palacios,stopa,martin}.   These   calculations
have been  carried out using a  fixed value of  the external parameter
$\omega_0$ of  the harmonic confining  potential. Our problem  here is
more complex because  for a given value of $a_1$  we need to determine
$a_2$  required to  satisfy Eq.~\ref{equal-mu}  and this  requires the
knowledge of  the full function  $E(N,a)$.  Next, we present  a simple
model to  express the  energy of  one dot in  a harmonic  potential of
external   frequency   $\omega_0(a)$.   We   represent   the   $N$-dot
wave-function  as a  Slater-determinant of  Hermite  polynomials which
correspond  to  a  2D  harmonic oscillator  potential  of  ``dressed''
frequency  $\omega$.  The value  of  $\omega$  will  be determined  by
minimizing the expectation value of the Hamiltonian which includes the
Coulomb  interaction. The  presence  of the  Coulomb interaction  will
decrease the value of $\omega$ compared to $\omega_0$. We find that
\begin{equation}
E(N,a) =  {1 \over 2}  \hbar (\omega + {{\omega_0^2}  \over {\omega}})
n(N) + u \sqrt {\hbar \omega} {{N(N-1)} \over 2}.
\label{energy}
\end{equation}
The values  of the  function $n(N)$ for  $N$ ranging from  $0$ through
$12$  are $0,1,2,4,6,8,10,13,16,19,22,25,28$  respectively.   The last
term corresponds to the  electron-electron interaction which scales as
$N(N-1)/2$  with  respect  to   $N$.  In  addition,  it  is  inversely
proportional to the dot size  which implies that it is proportional to
$\sqrt{\omega}$.   The parameter $u$  gives a  measure of  the Coulomb
interaction in the  dot when all the important  dependences are scaled
out and we expect it to  be almost independent of $\omega$ and $N$. In
the capacitance  model for large dots,  $u$ is a  constant. Assuming a
universal value  for $u$ for any  dot in a parabolic  potential, for a
given value of $\omega_0$ and $N$ the energy is minimized with respect
to $\omega$.   In the top part  of Fig.~\ref{dft}, the  results of our
calculation of  $\mu(N)$ are compared  with those of a  recent density
function theory (DFT) calculation\cite{martin} using $u=2~ meV^{-1/2}$
and the optimal value of $\omega$.  The agreement is very satisfactory
given the  fact that  the same value  of $u$  is used for  the results
obtained for three very different values of $\omega_0=4,10,20 meV$.

\section{Determination of the parameters of the quantum-dot array}
 
We take $\epsilon=12.9$ and $m^*/m =  0.067 $ for $GaAs$ and we choose
$N_1=6$, $N_2=12$,  $d=500 \AA$, $a_1=1000$ and $V_g=1~V$  and we find
that,  $\hbar  \omega_1   =  1.915  meV$  and  in   order  to  satisfy
Eq.~\ref{equal-mu}  we   need  to  take  $a_2=620   \AA$,  and  $\hbar
\omega_2=1.843   meV$.   The  left   and  the   right  hand   side  of
Eq.~\ref{equal-mu} are shown in the bottom of Fig.~\ref{dft}.  We note
that the  horizontal solid line  denotes that the two  functions share
the  same value  for $N_1=6$  and $N_2=12$.  Notice that  the chemical
potential    differences   $\delta    \mu(N)=    \mu(N)-\mu(N)$   are:
$1.69,1.64,4.46  meV$   for  the  larger  dot   and  $N=11,12,13$  and
$1.93,1.81,4.13 meV$ for the smaller dot and $N=5,6,7$.

We can tune the doping (controlled by the back-gate potential) to fill
the  two  dots with  $N_1=12$  and  $N_2=6$  electrons.  This  can  be
achieved by controlling  the total 2D electron density  to be $24/b^2$
and taking  $b=3400 \AA$ ($b$ should be  larger than $2(a_1+a_2)\simeq
3240 \AA$ for  our example here), this corresponds to  a 2D density of
$n=2.0~ 10^{10}/cm^2$.  If we reduce the value of the electron density
further we can  change the filling of the $p$-level  of the small dots
and of the $d$-level of the larger hole. By changing $V_g$, and $a_1$,
$a_2$ and $d$ we can increase the $\omega_0$'s which will increase the
required 2D electron density.

In  order to describe  the electron  hopping from  a $p$-level  of the
small dot to the $d_{x^2-y^2}$  level of the neighboring larger dot we
will introduce  the hopping matrix element $V_{pd}=\langle  p_x|_i H |
d_{x^2-y^2}  \rangle_{i+\hat  x}=   \langle  p_y|_i  H  |  d_{x^2-y^2}
\rangle_{i+\hat y}$.  Notice that  the hopping matrix elements between
$p$ and $d_{xy}$  of neighboring dots is zero,  while that between $p$
and  $d_{1-r^2}$ is  smaller and  can be  neglected  for appropriately
chosen  inter-dot  distances.   These  outer  electron  orbitals  with
significant  overlap integral are  shown in  Fig.~\ref{3dpot}.  Notice
the  direct   correspondence  of  the  orbitals  here   and  the  $Cu$
$d_{x^2-y^2}$ and  the oxygen $p_x$ and  $p_y$ in the  $Cu-O$ plane of
the copper-oxide materials. Furthermore, the orbital $d_{3z^2-r^2}$ of
$Cu$ in the $Cu-O$ plane,  whose role was much debated, corresponds to
the orbital $d_{1-r^2}$.

The tight binding Hamiltonian describing this quantum dot array is
\begin{eqnarray}
H = \sum_{i,l \in  (i),\sigma} ( \epsilon_d d^{\dagger}_{i\sigma} d_{i
 \sigma}   +  \epsilon_p   p^{\dagger}_{l  \sigma}   p_{l   \sigma}  +
 V_{pd}d^{\dagger}_{i\sigma}p_{l \sigma} + H.c.  )
\label{tbind}
\end{eqnarray} 
and  can be  analytically  diagonalized in  a straightforward  manner.
Here $\sum_{l\in (i)}$ denotes the sum over the neighbors of site $i$.
$d^{\dagger}_{i \sigma}$ and $p^{\dagger}_{l \sigma}$ create electrons
in   the  states   $|d_{x^2-y^2}>_i$  and   $|p_x>_l$   (or  $p_y>_l$)
respectively with  spin $\sigma$.  Large overlap can  be achieved when
the   inter-cell   distance   $b$   is   comparable   to   $b_o   \sim
4(\lambda_1^{-1/2} +  \lambda_2^{-1/2})$.  In the case  of the example
we  gave  above  $\lambda_1  \simeq  \lambda_2$  and  $b_o  \sim  2000
\AA$.  Taking   $\lambda_1=\lambda_2=\lambda$,  the  overlap  integral
between the $p$  and the $d$ states which are  separated by a distance
$b$ decays  as $e^{-(b/b_o)^2}$.  Thus, the magnitude  of $V_{pd}$ can
become  a large  fraction (of  the order  of 20\%)  of  the within-dot
electron kinetic energy $\hbar^2\lambda/2m^* \sim 1~meV$.

Based on  experimental results  on charging of  a quantum dot  using a
capacitor\cite{capacitor} and our results here one needs to include in
the Hamiltonian a term of the form
\begin{equation}
H_U     =      U_d     \sum_i     d^{\dagger}_{i\uparrow}d_{i\uparrow}
d^{\dagger}_{i\downarrow}d_{i\downarrow}+  U_p  \sum_{i,  l  \in  (i)}
p^{\dagger}_{l\uparrow}p_{l\uparrow}
p^{\dagger}_{l\downarrow}p_{l\downarrow}.
\label{hubbard}
\end{equation}
$U_{d,p}$ are of the order $\delta\mu_1(12)$ or $\delta \mu_2(6)$ both
of  which  are  $\sim  2~meV$.   Including  this  term  in  the  above
tight-binding Hamiltonian  we obtain  the same two-band  Hubbard model
which has been used to describe a single layer of copper-oxide.  Short
as well  as long range Coulomb  interaction should be  included in the
above Hamiltonian in order to  understand the phase diagram of such an
array of  quantum dots.  However,  even when one attempts  to describe
the physics  of the $Cu-O$ planes in  the copper-oxide superconductors
one should include these interactions.

\section{Stripe Formation on the array and  their Imaging}

The copper-oxide materials exhibit\cite{Tranq}  stripe formation 
at filling factor
of $1/8$.   Numerical
studies\cite{emery,hellberg2,hellberg1}  of the t-J  model, which is a
possible  reduction  of   the  above  three-orbital  model\cite{zhang},
indicate  that the  model has  a phase  separation  instability.  Some
different numerical  studies\cite{white1} of this  model indicate that
the system at least with cylindrical boundary conditions seems to form
stripes   at   the   appropriate   filling   factor   and   value   of
$J/t$.  Independently   of  the  controversy   surrounding  the  stripe
formation in the t-J  model\cite{hellberg1,white2}, it may be expected
that stripes form when the  long range part of the Coulomb interaction
is included in the t-J model.
At the appropriate filling factor one might expect
formation of  stripes in the quantum dot  system described previously.
The  quantum  dot array  discussed  here  cannot  allow a  macroscopic
electronic charge separation,  thus, we expect (just like  in the real
copper-oxide superconductors)  to see a striped state  or another form
of clustering  of charge and spin.          

The  formation of  the  stripe  state can  be  investigated using  the
quantum dot array  proposed in the present paper.   We expect that the
stripe state can be detected  by electric force microscopy (EFM).  
Since the inter-dot distance  is of the order of  $0.1 \mu m$,
atomic  scale spatial  resolution is not required.  However,  
we  need to  be able  to
detect electric potential variation on  the surface of the quantum dot
array of the  order of $1 mV$ or smaller.  Therefore a special purpose
tip coated  with a metal layer  should be made which  should be wider
than the typical size in order to detect the voltage change associated
with  such  inhomogeneous charge  distribution  when stripe  formation
occurs.

Transport measurements can be  also performed which can possibly shed light
on the  original problem.  Conductance measurements have  already been
performed  on  one-dimensional  arrays  of  quantum  dots\cite{kouwenhoven}. 
 There  are
indications\cite{sarma2}   that   an   Anderson-Mott   metal-insulator
transition might have been observed\cite{kouwenhoven,haug}. At filling
factors  around  one  hole per  unit  cell,  one  expects to  find  an
antiferromagnetically ordered insulator.  By applying a magnetic field
the structure can  be converted to a metal.   In addition, capacitance
measurements can be used to  measure the ``addition'' energy of adding
an extra electron to the quantum dot array.

We would  like to discuss  decoherence or the  effect of noise  on the
proposed device.  In a universal quantum computer not only the devices
representing q-bits are required to be free from noise and decoherence
but,  in addition, one  needs to  be able  to manipulate  them without
destruction of coherence. The latter issue obviously does not arise in
our  case of the  dedicated quantum  simulator.  The  proposed device,
however,  and the  original system  are both  affected  by decoherence
effects. The CuO planes are hardly  in a vacuum, phonons being not far
in energy  from the energy  separation between the  electronic states.
What is being  proposed is a device for  simulating a particular model
Hamiltonian, which  may have many  features in common with  2D lattice
models.  So the  question is to what extent one  can neglect noise and
get reliable answers  concerning mostly thermodynamic information that
is discussed in this paper.  One expects that there will be regions of
the thermodynamic phase  diagram of the proposed device  which will be
strongly affected  by impurities,  imperfections and other  sources of
noise  and regions  which will  not  be strongly  affected.  The  full
calculation of the effects of such noise on the phase diagram of these
models is left as an open problem; to answer it one needs to carry out
a  more  complete calculation  of  the  electronic  properties of  the
proposed device and their sensitivity to the various sources of noise.

In summary,  we have shown that  the 2D quantum-dot  array produced by
the structure  shown in  Fig.~\ref{hetero2} using the  calculated hole
sizes  and proposed  gate voltages  and doping  values, maps  onto the
strong    coupling    limit    of    the    Hamiltonian    given    by
Eqs.  (\ref{tbind},\ref{hubbard}). This  Hamiltonian is  that  used to
describe  the  physics  of   the  $Cu-O$  plane  of  the  copper-oxide
superconductors. This quantum dot array can be considered as
an ``analog''  quantum computer (as opposed to digital) or a
dedicated quantum simulator of the dynamics of the $Cu-O$ planes
in these materials. In particular we have shown that 
the striped state formed in the above materials, which seems to be hard
to study experimentally\cite{Tranq,Mook}, can be studied and analyzed
using an analog system of such a 2D quantum dot array 
which is predicted to form a stripe state with a wavelength of mesoscopic 
size. This allows conventional imaging methods such as EFM 
to be used.

\section{Acknowledgements}

This work was supported in part by an Office of Naval Research grant
number N00014-99-1-1094. The author wishes to thank S. Von Molnar and 
P. Xing for useful discussions.

\end{document}